\documentclass[12pt,preprint]{aastex}

\shorttitle{Turbulence in Local Clouds}
\shortauthors{Spangler et al}

\begin{document}

\title{Properties of Turbulence in the Very Local Interstellar Clouds}
\author{Steven R. Spangler and Allison H. Savage}
\affil{Department of Physics and Astronomy, University of Iowa, Iowa City, IA 52242}
\author{Seth Redfield}
\affil{Department of Astronomy, Wesleyan University, Van Vleck Observatory, Middletown, CT 06459}
\begin{abstract}
We have investigated the degree to which turbulence in the Very Local Interstellar Clouds resembles the highly-studied turbulence in the solar corona and the solar wind. The turbulence diagnostics for the Local Clouds are the absorption line widths measured along 32 lines of sight to nearby stars, yielding measurements for 53 absorption components \citep{Redfield04}.  We have tested whether the Local Cloud turbulence has the following properties of turbulence in the solar corona or the solar wind: (a) velocity fluctuations mainly perpendicular to the average magnetic field, (b) a temperature anisotropy in the sense that the perpendicular temperature is larger than the parallel temperature (or at least enhanced relative to expectation), and (c) an ion temperature which is dependent on the ion Larmor radius, in the sense that more massive ions have higher temperatures.  Our analysis of the data does not show compelling evidence for any of these properties in Local Cloud turbulence, indicating possible differences with heliospheric plasmas. In the case of anisotropy of velocity fluctuations, although the expected observational signature is not seen, we cannot exclude the possibility of relatively high degrees of anisotropy (anisotropy parameter $\epsilon \sim 0.50 - 0.70$), if some other process in the the Local Clouds is causing variations in the turbulent line width from one line of sight to another.  We briefly consider possible reasons for differences between coronal and solar wind turbulence and that in the Local Clouds. The apparent absence of anisotropy of the velocity fluctuations and ion temperature might be due to randomization of the interstellar magnetic field on spatial scales $\sim$ 10 parsecs, but this would not explain the absence of ion mass-dependence in the ion temperature. A likely explanation of all the results is the greater collisionality, due to ion-neutral collisions, of the partially-ionized Local Cloud plasma.   

\end{abstract}


\section{Introduction}
Turbulence is believed to be an extremely important phenomenon in many astrophysical media, such as the diffuse phases of the interstellar medium, molecular clouds, supernova remnants, accretion disks around compact objects, extragalactic radio sources, and the intracluster media of clusters of galaxies.  It is speculated that this turbulence may serve such functions as providing additional sources of pressure or heat input, determining transport coefficients such as viscosity and resistivity, and accelerating charged particles to high energy.  Assessment of these possibilities requires knowledge of the properties of the turbulence in some detail.  To address these possible roles of turbulence, we need to know not only the power spectrum, which may be nearly identical for turbulent excitations in vastly different cases \citep[see, for example, ][]{Bayley92}, but more importantly, the symmetries and relationships between fluctuations in plasma velocity, magnetic field, density, etc.  

Such knowledge is almost unattainable for most media of interest to astronomers.  In many astrophysical plasmas, even the average plasma parameters are incompletely known or totally unknown. Information on the properties of turbulence is almost always highly indirect in the spatially-averaged, path-integrated measurements available to astronomers.  Given this, a-priori knowledge of the nature of turbulence from independent sources becomes very important. 

Observations of the magnetohydrodynamic (MHD) turbulence which exists in the solar wind play a very important role in our understanding of turbulence, and solar wind data sets have been the primary data sources in the development and validation of theories of MHD turbulence. In this paper, we consider the properties of turbulence in another medium which is, perhaps surprisingly, very well diagnosed. This medium is the partially-ionized plasma contained in clouds in the Very Local Interstellar Medium (VLISM) within about 15 parsecs of the Sun.  The properties of these clouds, and the ways in which turbulence in them is measured, are discussed in Section 3. The goal of this paper is to investigate the extent to which the turbulence in these clouds resembles, or differs from, the turbulence in the solar corona and the solar wind. 

A preliminary report on this topic was given in \cite{Spangler10}. That paper pointed out that the high spectral resolution absorption line measurements of \cite{Redfield01} and \cite{Redfield04} could be used to extract properties of turbulence in the Local Clouds, and that the inferred properties could be compared with those of the solar corona and solar wind.  \cite{Spangler10} drew preliminary conclusions from examination of the results published in \cite{Redfield04}.  In the present paper, we use detailed and quantitative analyses of the data of \cite{Redfield04} to study the plasma turbulence in the Local Clouds. 

\section{Properties of Solar Wind Turbulence}
The obvious advantage of  solar wind turbulence is that basic plasma physics measurements of vector magnetic field, plasma flow velocity, density, temperatures, and even electron and ion distribution functions can be measured in situ with spacecraft.  The fluctuations in all of these quantities have been extensively studied, a large literature written, and major conclusions reached. Among the many influential articles and reviews of the subject are \cite{Bavassano82}, the monograph by \cite{Tu95}, and the review articles by \cite{Goldstein95} and \cite{Bruno05}.  

Another nearby plasma with extensive diagnostics (although not, as yet, in situ measurements) is the solar corona.  Our knowledge of the corona and its turbulence results from high spatial resolution images, ultraviolet spectroscopy of numerous transitions, and radio propagation measurements. In addition, a sort of ``ground truth'' of coronal plasma measurements is provided by spacecraft measurements at heliocentric distances from 0.28 to 1 astronomical units (AU).  The coronal plasma is convected out into space and becomes the solar wind.   Among the many reviews of the coronal plasma, two which are particularly relevant to the present investigation are \cite{Cranmer02} and \cite{Bird90}. 

A list of the main properties of solar wind and coronal turbulence could be extensive.  We list four properties which are particularly relevant to the present investigation. 
\begin{enumerate}
\item The fluctuations in magnetic field and plasma flow velocity are highly correlated.  The equations of magnetohydrodynamics couple fluctuations in magnetic field and plasma flow velocity. The dimensionless amplitude of the velocity fluctuations, $\delta v/V_A$ is highly correlated with the dimensionless amplitude of the magnetic field fluctuations, $\delta b/B_0$, where $\delta b$ and $\delta v$ are the root-mean-square (rms) fluctuations in magnetic field and flow velocity, respectively, and $V_A$ and $B_0$ are the Alfv\'{e}n speed and magnitude of the magnetic field \citep{Spangler04}.  This property can be used to infer the magnitude of fluctuations in one of the quantities ($\delta b$ or $\delta v$), given a measurement of the other. 
\item The fluctuations in magnetic field and velocity are predominantly perpendicular to the large scale interplanetary magnetic field.  This property can be readily understood if the solar wind turbulence is viewed as an ensemble of interacting Alfv\'{e}n waves, or if it is described by the equations of quasi-2D magnetohydrodynamics \citep{Zank92}. Observations illustrating this property are presented in \cite{Bavassano82} and \cite{Klein93}. \cite{Bavassano82} and \cite{Klein93} also show that transverse velocity fluctuations are not a universal property of solar wind turbulence.  They find that the anisotropy of the turbulence (i.e. the excess of perpendicular over parallel fluctuations) decreases with heliocentric distance, and is less pronounced in the slow speed solar wind relative to the high speed wind. 
\item A plasma heating process is occurring which preferentially increases the perpendicular (to the magnetic field) temperature relative to the parallel temperature. This is a hallmark of ion heating by ion cyclotron resonance mechanisms. In the case of the corona, the perpendicular temperature $T_{\perp}$ exceeds the parallel temperature $T_{\parallel}$ by a large factor \citep{Cranmer02, Hollweg08}.  In the case of the solar wind, typically $T_{\perp} \leq T_{\parallel}$, but the observed values are much higher than would be expected in an expanding solar wind without preferential perpendicular heating \citep{Kasper09}.  This is often expressed as a systematic increase of the first adiabatic invariant with heliocentric distance in the solar wind.  \cite{Chandran10} has recently proposed that another heating mechanism termed ``stochastic acceleration'', due to significant velocity fluctuations on scales comparable to the ion cyclotron radius, can also produce the observed features of anisotropic temperatures.  Chandran's mechanism does not require a cyclotron resonance between ions and plane Alfv\'{e}n waves. Regardless of the correct theoretical explanation, the aforementioned phenomena are most pronounced for the solar corona and the high speed solar wind, i.e. the most collisionless parts of the heliosphere. Finally, it is worth noting that in collisionless heliospheric plasmas, the ion and electron distribution functions are not Maxwellians \citep[e.g.][]{Tu95}, a fact which may be of great significance to the thermodynamics of these plasmas \citep{Scudder92}.    
\item Different ions, characterized by different Larmor radii, have different temperatures.  Those ions with larger Larmor radii have higher temperatures. This property, which is also readily understood on the basis of ion cyclotron resonance mechanisms \citep{Hollweg08} as well as the stochastic acceleration mechanism of \cite{Chandran10}, is highly pronounced in the solar corona, where the spectroscopically-inferred temperature of OVI is at least 30 times the proton temperature \citep{Cranmer02}. A particularly striking illustration of the observed variation of temperature with ion species is given in \cite{Moran03}, particularly Figure 5 of that paper.  
\end{enumerate} 
These properties are prominent in the highly collisionless plasmas of the corona and the high speed solar wind, and become less pronounced in the denser, slower, and more collisional parts of the slow speed solar wind \citep{Kasper09}.  
It is of great interest to ask whether other astrophysical plasmas also possess turbulence with these properties. A positive answer would permit immediate application of the considerable body of knowledge and understanding of heliospheric turbulence to the numerous media listed at the beginning of this section. The specific goal of this paper is a modest step in that direction. We explore the extent to which turbulence in the Local Clouds of the VLISM resembles the turbulence in the solar wind and solar corona.  

\section{The Clouds of the Very Local Interstellar Medium (VLISM) and the Turbulence They Contain} 
The Very Local Interstellar Medium (VLISM) is loosely defined as the interstellar medium within about 15 parsecs of the Sun.  One of the interesting aspects of the VLISM is that it contains about 15 clouds with diameters of a few parsecs \citep{Redfield08a}. It appears that the Sun is near the interface and region of interaction of two of these clouds,  the Local Interstellar Cloud, or LIC, and the G cloud \citep{Redfield08a}. Reviews of the properties of these clouds may be found in \cite{Frisch00}, \cite{Redfield09}, and \cite{Frisch11}. Most of the information we have about these clouds comes from UV and visible wavelength spectroscopy.  Absorption lines attributable to these clouds are measured along lines of sight to nearby stars with precisely known distances. Properties of these clouds are deduced from the Doppler shift, strength, and width of the spectral lines. These clouds are plasmas because absorption lines of ions as well as neutral atoms are observed; the ionization fraction is about 50 \% \citep{Redfield08b}.  

Although the information available on these clouds is not as extensive as for the solar wind or solar corona, it is sufficient to place the Local Clouds among the best-diagnosed astrophysical plasmas.  There are several reasons for this state of affairs. First, because the absorption lines are measured in the spectra of nearby stars with precisely known distances, the spatial extent of the clouds is well determined.  Second, the neutral component of the clouds flows into the inner solar system, where it can be measured in situ \citep[e.g.][]{Moebius09}. Finally, the heliosphere is embedded in one of these clouds, the LIC cloud, and the solar wind interacts with it.  The shape and other characteristics of the solar wind interaction provide constraints on the LIC cloud properties \citep{Lallement05,Opher09}. 
The mean plasma properties of the turbulent clouds are given in Table 1 \citep[adapted from][]{Redfield08a,Redfield08b}. 
\begin{deluxetable}{cr}
\tabletypesize{\small}
\tablecaption{Mean Plasma Parameters of Local Interstellar Clouds\label{tbl-1}}
\tablewidth{0pt}
\tablehead{\colhead{Plasma Parameter} & \colhead{Value}}
\startdata
electron density & 0.11 cm$^{-3}$ \\
neutral density &  0.1 cm$^{-3}$  \\
temperature &  4000-8000 K (typical) \\
magnetic field & 3-4 $\mu$ G (assumed)
\enddata
\end{deluxetable}  

Information on turbulence in the Local Clouds is discussed in \cite{Redfield04}.  Such information is retrievable because the absorption line width $b$ can be measured for transitions of several atoms or ions.  \cite{Redfield04} fit the line width data for each line of sight and Doppler component to the formula
\begin{equation}
b^2 = \frac{2 k_B T}{m} + \xi^2
\end{equation}
where $T$ is the temperature (assumed the same for all atomic and ion species), $k_B$ is Boltzmann's constant, $m$ is the mass of the atom or ion, and $\xi$ is the non-thermal Doppler width of the line, attributable to turbulent flows in which all ions and atoms participate.  

To anticipate one of the main points of this paper, a coronal astronomer or solar wind physicist would immediately take issue with Equation (1), noting point (4) above that in those media, different atoms and ions have different temperatures.  In Section 4.4 we will investigate the degree to which a single, common temperature characterizes the Local Clouds.  
\section{A Study of Local Cloud Turbulence Properties} 
In this section, we test whether the four properties of solar wind turbulence stated in Section 2 above also characterize the turbulence in the Local Clouds.  The data set we use are published measurements of $T$ and $\xi$ given in \cite{Redfield04} and \cite{Redfield01}.  We also have used the line width measurements $b$ for different atoms and ions.  These data are shown in Figure 1 of \cite{Redfield04}.  We have used the numerical versions of those data, which are available for many lines of sight. 

Data on $T$ and $\xi$ are given in \cite{Redfield04} and \cite{Redfield01} for 32 lines of sight, possessing 53 absorption line components. Data on 50 absorption components are given in Table 1 of \cite{Redfield04}.  Data for the remaining 3 components are given in Table 5 of \cite{Redfield01}. Since some of the 32 lines of sight intercept more than one cloud, we have a larger number of absorption components than lines of sight. Each of the 53 components provides an independent observational estimate of the properties of turbulence in one of the 15 Local Clouds.  
In the following, we analyse these data and investigate the degree to which the Local Cloud turbulence adheres to the characteristics listed in Section 2. 
\subsection{Velocity Fluctuations Perpendicular to $\vec{B}_0$}
We interpret the turbulent line width parameter $\xi$ to be a measure of the velocity fluctuations in the cloud turbulence.  If solar wind turbulence is a good model for the cloud turbulence, this turbulence is Alfv\'{e}nic, and the velocity fluctuations should be perpendicular to the large scale interstellar magnetic field $\vec{B}_0$.  If this is the case, the measured value of $\xi$ should vary with position on the sky. 

The reasoning behind this statement is illustrated in Figure 1. 
\begin{figure}[h]
\epsscale{0.60}
\includegraphics[width=18pc]{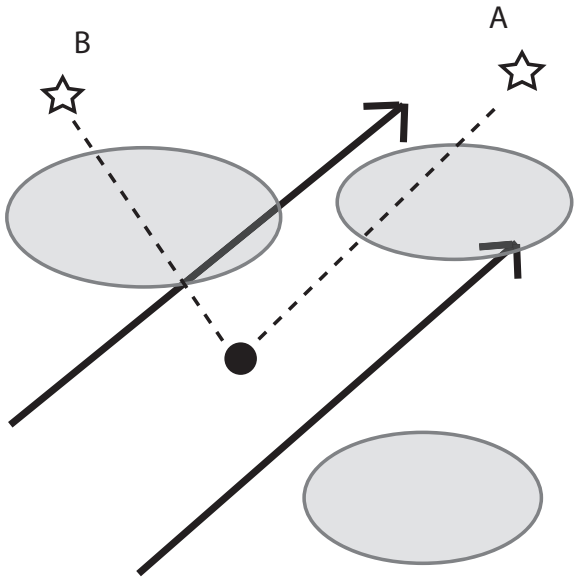}
\caption{An illustrated argument why the turbulent broadening parameter $\xi$ should depend on direction on the sky, if the turbulence is Alfv\'{e}nic with velocity fluctuations $\perp$ to the large scale interstellar magnetic field $\vec{B}_0$.  The heavy arrows indicate the direction of $\vec{B}_0$, and the shaded regions indicate the relatively dense Local Clouds, embedded in the rarefied Local Cavity. A line of sight along the large-scale field (to star A) will show little turbulent line broadening because the fluctuations are perpendicular to the line of sight.  A line of sight across the field (to star B) will show large turbulent line broadening because the turbulent velocity fluctuations are aligned with the line of sight.  Figure taken from \cite{Spangler10}.}
\end{figure}
In certain directions on the sky, we are looking across $\vec{B}_0$, and the turbulent fluctuations should be along the line of sight (more properly, one of the two fluctuating magnetic field components in a plane perpendicular to $\vec{B}_0$ will be aligned with the line of sight).  For other directions, we are looking along $\vec{B}_0$, and the turbulent velocity fluctuations are mainly transverse to the line of sight.  In this case, $\xi$ should be small.  

The direction of $\vec{B}_0$ must be considered an unknown parameter. Although there is information on the form of the global Galactic magnetic field from Faraday rotation measurements of pulsars and extragalactic radio sources \citep{Rand89,Rand94,Minter96,VanEck11} as well as measurements of the polarized Galactic synchrotron emission \citep{Beck96,Haverkorn04a}, the magnetic field models are deduced from measurements on lines of sight which are kiloparsecs in length.  All analyses of the galactic magnetic field agree that the fluctuating (presumably turbulent) component of the galactic magnetic field is comparable to or larger than the systematic component \citep[e.g.][]{Rand89,Minter96,Haverkorn04a,VanEck11}. By systematic component, we mean a vector field which is describable by a relatively simple function of Galactocentric coordinates. 

In a very local sense, the ``large scale'' magnetic field is almost certainly dominated by these turbulent fluctuations.  It may reflect the random orientation of the largest eddy in the solar neighborhood.  In any case, $\vec{B}_0$, and the unit vector in the direction of $\vec{B}_0$, $\hat{b}$, can point in any direction in the sky.  

Before proceeding further, it is necessary to note that there are two independent, and incompatible estimates of the direction of $\hat{b}$. \cite{Lallement05} use the difference between the direction of neutral helium flow and that of the largest concentration of neutral hydrogen outside the heliopause\footnote{The heliopause is the contact discontinuity between the shocked solar wind and the interstellar medium.} to infer that $\hat{b}$ points in the range $l = 205^{\circ} - 240^{\circ}$, $b = -60^{\circ} - -38^{\circ}$.  \cite{Gurnett06} report measurements of the direction to sources of low frequency radio emission, which are assumed to be generated in the heliosheath \footnote{The heliosheath is the region of shocked solar wind between the solar wind termination shock and the heliopause.}.  They assume that this radio emission is generated at points on the heliopause which are perpendicular to $\vec{B}_0$.  They do not retrieve the vector $\hat{b}$, but report the angle between $\hat{b}$ and the direction to the ecliptic pole.  They interpret their results as being consistent with \cite{Lallement05}.  

A model-dependent estimate of $\hat{b}$ has also been presented by \cite{Opher09}.  \cite{Opher09} use Voyager 1 measurements of the plasma flow direction in the heliosheath, together with an MHD model of the heliosheath, to infer that $\hat{b}$ points in the approximate direction $l = 10^{\circ} - 20^{\circ}$, $b = 28^{\circ} - 38^{\circ}$.  The estimates of \cite{Lallement05} and \cite{Opher09} appear to be significantly different.  It is worth emphasizing that all of the aforementioned techniques are model-dependent in that they adopt physical assumptions about processes in the outer heliosphere, or use MHD simulations of the heliosphere to relate the actual measured quantity to the properties of the solar wind-ISM interaction, including the direction of $\hat{b}$.  

To carry out an analysis suggested by Figure 1, we need a model for the form of the turbulent line width $\xi$ in the case of anisotropic, Alfv\'{e}nic turbulence in the Local Clouds.  The derivation of such an expression is given in the Appendix.  We assume the turbulence is characterized by a root-mean-squared amplitude of the perpendicular velocity fluctuations $V_{\perp}$, and an anisotropy factor $\epsilon$.  When $\epsilon$ is zero, the turbulence is isotropic, and if $\epsilon = 1$ the turbulent motions lie completely in the plane perpendicular to $\vec{B}_0$, with no motion in the direction of the field.  In this case, (see Appendix for derivation)
\begin{equation}
\frac{\xi^2}{V_{\perp}^2} = \frac{<v_L^2>}{V_{\perp}^2} = 1 - \epsilon (\sin b \sin \beta + \cos \Delta l \cos b \cos \beta )^2
\end{equation}
In addition to variables already defined, the galactic coordinates of the direction of the line of sight are $(l,b)$ and the galactic coordinates defining the direction of the local magnetic field are $(\lambda, \beta)$. The angle $\Delta l \equiv \lambda - l$. The component of the gas velocity along the line of sight is given by $v_L$. Equation (2) is useful for analyses of the sort to be described shortly but a more intuitive expression is 
\begin{equation}
\frac{\xi^2}{V_{\perp}^2} = \frac{<v_L^2>}{V_{\perp}^2} = 1 - \epsilon \cos^2A
\end{equation}
where $A$ is the angle between the line of sight and the local field.  

We cannot, in a straightforward way, test whether actual data adhere to the relationship in Equation (3) without knowledge of the direction of the local magnetic field, indicated by the unit vector $\hat{b}$ or the angles $(\lambda, \beta)$.  We therefore adopted the following, brute-force approach, which was undertaken without reference to the a-priori estimates of the magnetic field direction proposed by \cite{Lallement05} and \cite{Opher09}.  
\begin{enumerate}
\item We chose 361 candidate directions for $\hat{b}$, each characterized by  values of $(\lambda, \beta)$.  These candidate directions were spaced 10 degrees apart in galactic latitude and longitude, and completely covered one hemisphere of the sky.  Given the nature of the anisotropy sought, one hemisphere is adequate for complete coverage.  
\item For each candidate direction, the angle $A$ could be calculated (using Equation (2)) for each line of sight for which estimates of $T$ and $\xi$ are available from \cite{Redfield04} and \cite{Redfield01}.  We then made a plot of $\frac{\xi^2}{V_{\perp}^2}$ versus $\cos A$ for this candidate direction of $\hat{b}$.  
\item Each plot was examined to see if a relationship of the form given in Equation (3) could be discerned.  
\end{enumerate}

No compelling cases for such a relationship were found.  A set of cases in which something like the expected relationship seemed to be present in the data (often referred to by self-deluded individuals as ``tantalizing'') were collected for further scrutiny. In these cases, there seemed to be a larger average value of $\xi^2$ for smaller values of $|\cos A |$ than for larger values. The total number of such cases was 10, and these cases were roughly clustered in the direction $\lambda \sim 40^{\circ}, \beta \sim 60^{\circ}$. The case for $(\lambda=40^{\circ}, \beta=60^{\circ})$ is shown in Figure 2. 
\begin{figure}[h]
\epsscale{0.60}
\includegraphics[width=22pc,angle=-90]{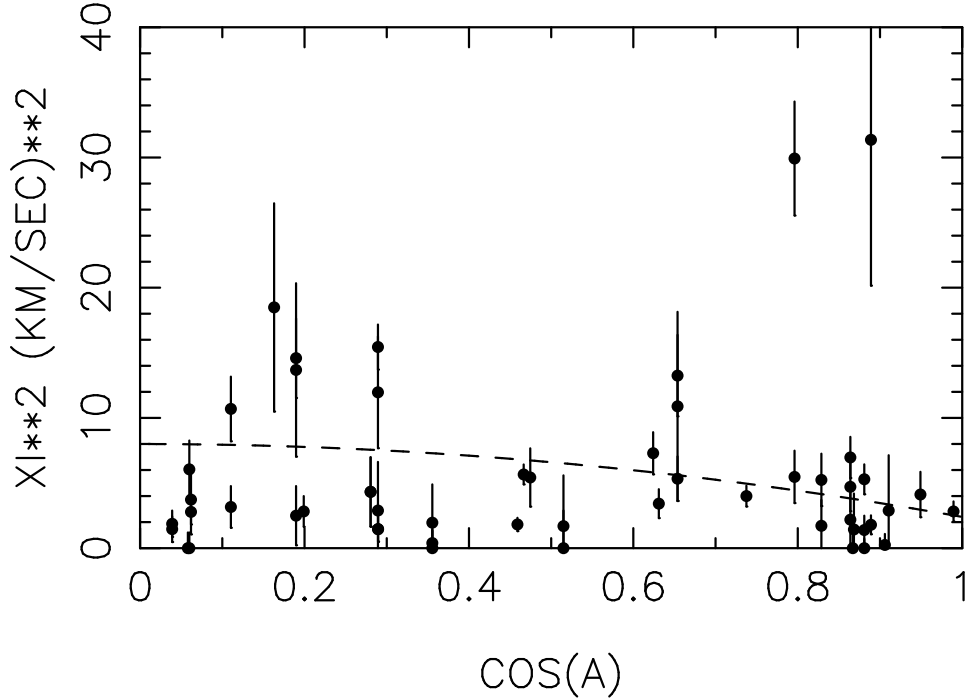}
\caption{Plot of $\xi^2$ versus $| \cos A |$, when the  local interstellar magnetic field is assumed to point in the direction $\lambda =40^{\circ}, \beta = 60^{\circ}$. The plot shows measurements of $\xi^2$ extending to larger values for $\cos A \leq 0.40$, as expected for transverse velocity fluctuations. The statistical significance of this difference is discussed in Section 4.1.1. The dashed line shows the model given by Equation (3) with $V_{\perp}^2=8.00$ km$^2$/sec$^2$, and $\epsilon = 0.70$. }
\end{figure}
We do not claim this result as a detection of anisotropy in the Local Cloud turbulence, but it does give an indication of the data quality in one of the best cases for anisotropy. An obvious feature in Figure 2 is the presence of two data points at $| \cos A | \simeq 0.8 - 0.9$ with possibly anomalous values of $\xi^2$. These points correspond to the +2.6 km/sec radial velocity component for Alkaid (HD 120315) and the +13.9 km/sec radial velocity component for $\zeta$ Doradi (HD 33262).  Obviously, any hint of a systematic dependence of $\xi^2$ on $A$ disappears if these are valid points.  However, it appears that they lie well outside the distribution of $\xi^2$ values for stars with similar values of $\cos A$.  There is basis for suspecting that the line widths might be affected by blends of two or more components, thus inflating a single component fit to the line.  For this reason, Alkaid and $\zeta$ Dor may be provisionally considered outliers. In the analyses which follow, we have considered the complete data set of 53 absorption measurements, as well as an edited subset in which Alkaid and $\zeta$ Doradi are removed.  
\subsubsection{A Search for Weaker Anisotropy and Quantitative Limits to the Anisotropy} 
Figure 2 does not present a strong case for anisotropy, defined as close adherence of the data to the expression given in Equation (3). However, it is possible that an anisotropy of the sort we are seeking is present, but obscured by star-to-star variations of another, unknown nature.  To detect anisotropy in this case, it is necessary to average measurements for several stars. Furthermore, we need a means of extracting from the data a quantitative upper limit to the anisotropy parameter $\epsilon$.   

A simple way of addressing both of these points is to average the data over intervals in $\cos A$.  For these purposes, we consider the star-to-star variations as noise superposed to a true signal of the form in Equation (3).  If anisotropy is present, the average value of $\frac{\xi^2}{V_{\perp}^2}$ for all lines of sight with $0 \leq \cos A \leq 0.3$  will be larger than for all lines of sight with $0.7 \leq \cos A \leq 1.0$.  Furthermore, the ratio of the mean values for $\xi^2$ in the two ranges of $\cos A$ is a measure of, or upper limit to, the anisotropy. 
The following analysis was undertaken. 
\begin{enumerate}
\item For each of the 10 lines of sight for which there was some suggestion of anisotropy, as in the case of Figure 2, we computed a list of $\xi^2$ versus $\cos A$.  Once again, we point out that an assumed, candidate direction for the local interstellar magnetic field is necessary to calculate the angle $A$.  
\item The mean value of $\xi^2$ was calculated for all measurements in two ranges in $\cos A$: $0 \leq \cos A \leq 0.3$, and $0.7 \leq \cos A \leq 1.0$, as well as an estimate of the error in the mean. The error in the mean of  $\xi^2$ in the intervals was calculated as follows.  We used the measured mean and standard deviation of $\xi$ values (2.24 and 1.03 km/sec respectively) given by \cite{Redfield04}, in Figure 2 of that paper. These values were used to compute the mean and standard deviation of the quantity $\xi^2$.  Finally, the standard deviation of the mean of the quantity $\xi^2$ for a sample of 16 measurements (30 \% of 53 data points) was calculated. 
\item A ratio $R$ was calculated in which the numerator was $\bar{\xi^2} \equiv <\xi^2>_1$ in the first interval, and the denominator was $\bar{\xi^2} \equiv <\xi^2>_2$ in the second interval.  The error in this ratio was calculated in the standard way, using the standard deviation of the mean of $\xi^2$ described in item \# 2 above.  Our value for the standard deviation in $R(0.3)$, used for all directions, was 0.33.  
\item Steps \# 2 and \# 3 were repeated for the broader intervals of $0 \leq \cos A \leq 0.5$ and $0.5 \leq \cos A \leq 1.0$ ($a = 0.5$). The associated error in $R(a=0.5)$ was taken to be 0.26.  
\end{enumerate}

The reason for carrying out the ratio analysis for two values of the interval width a, $a=0.3$ and $a=0.5$, is as follows. An analysis of this sort has competing demands on the value adopted for the width of the averaging interval, $a$. The smaller the value of $a$, the greater will be the contrast between the mean values of $<\xi^2>$ for the two intervals.  On the other hand, a larger value of $a$ results in more stars and absorption components in each bin, and thus a statistically more stable value of $<\xi^2>$.  We carried out an analysis for both $a=0.3$, to try and get the largest possible contrast with a significant number of data points contributing to the average, as well as $a=0.5$, which has lower contrast but includes all the data in the sample. 

The mean value of $\xi^2$ in each of the intervals, and their ratio, is easy to calculate from the data. To relate this ratio to the anisotropy factor $\epsilon$ requires use of Equation (3) for the expected relationship $\xi^2(A)$. Given Equation (3), the mean value $<\xi^2>_1$in the first interval $0 \leq \cos A \leq a$ is given by
\begin{equation}
<\xi^2>_1 = \frac{V_{\perp}^2}{a} \int_0^a (1 - \epsilon x^2) dx
\end{equation}
and similarly with the second interval, so the ratio of the two means, $R(a)$ is given by
\begin{equation}
R(a) \equiv \frac{<\xi^2>_1}{<\xi^2>_2} = \frac{\int_0^a (1 - \epsilon x^2) dx}{\int_{1-a}^1 (1 - \epsilon x^2) dx} = \frac{1 - \epsilon a^2/3}{1 - \epsilon (a^2 -3a +3)/3} = R(a,\epsilon)
\end{equation}
A plot of $R(a, \epsilon)$ for $a=0.3 \mbox{ and } a=0.5$ is shown in Figure 3. 
\begin{figure}[h]
\epsscale{0.60}
\includegraphics[width=18pc]{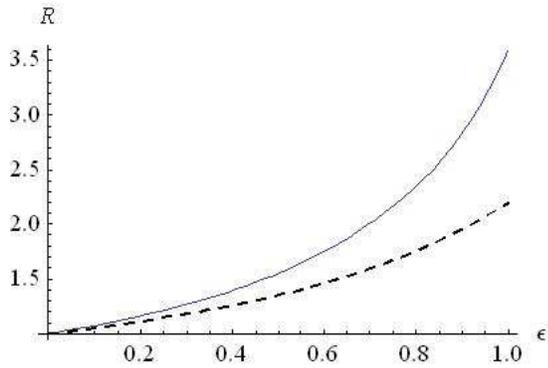}
\caption{Plot of the ratio $R(a,\epsilon)$ for $a=0.3$ (solid line) and $a=0.5$ (dashed line).  Given an interval for averaging data ($a=0.3, 0.5$), the measured ratio of the mean values of $\xi^2$ in the intervals $0 \leq \cos A \leq a$ and $1-a \leq \cos A \leq 1$ corresponds to the ordinate, and the inferred value of the anisotropy factor $\epsilon$ is the abscissa.   }
\end{figure}

Values of $R(a=0.3)$ and $R(a=0.5)$ and associated anisotropy indices $\epsilon$ were calculated for the 10 candidate directions for $\hat{b}$ described in Section 4.1.  Data from all 53 lines of sight were used.  In only one case, for $\lambda=10^{\circ}, \beta=40^{\circ}$, did $R(a=0.3)$ exceed unity by more than twice the adopted error ($R=1.71 \pm 0.33$), and even in this case the value of $R(a=0.5)$ was consistent with unity.  For the other nine directions, both $R(0.3)$ and $R(0.5)$ were consistent with unity (i.e. within 2 standard deviations of unity).  Of the twenty calculated quantities ($R(0.3)$ and $R(0.5)$ for 10 directions), half had $R \leq 1$, which is inconsistent with a velocity anisotropy of the sort we are seeking, but is consistent with random variations in the case when $\xi^2$ is isotropic.

The results of this analysis may be summarized as follows.  If the data for Alkaid and $\zeta$ Doradi are valid values, then there is no candidate magnetic field direction with a statistically significant value of $R > 1$, and corresponding value of $\epsilon$ different from 0.  

We now repeat this analysis, but assuming that Alkaid and $\zeta$ Doradi are outliers which may be excluded from the sample.  The results are presented in Table 2. The first four columns of this table contain, respectively, the galactic latitude and longitude of the candidate field direction, the measured value of $R$ with $a=0.3$ and associated error, and the value of the anisotropy parameter $\epsilon$ consistent with $R$ and its error, obtained from Equation (5).  The final two columns give the value of $R$ with $a=0.5$ and the value of the anisotropy corresponding to this value of $R$. The first ten rows correspond to the directions selected from our visual examination of plots similar to Figure 2.  The candidate field directions in the bottom two rows correspond to those proposed by  \cite{Opher09} and \cite{Lallement05}, respectively, and are discussed further in Section 4.3.   For the moment, we restrict consideration to the first 10 rows of Table 2, which were directions $(\lambda,\beta)$ chosen by us for closer examination. 
\begin{deluxetable}{rrrrrr}
\tabletypesize{\small}
\tablecaption{Averaging and Anisotropy Analysis-Alkaid and $\xi$ Doradi Excluded}
\tablewidth{0pt}
\tablehead{\colhead{$\lambda$} & \colhead{$\beta$} & \colhead{$R(0.3,\epsilon)$} & \colhead{$\epsilon_1$} & \colhead{$R(0.5,\epsilon)$} & \colhead{$\epsilon_2$}}
\startdata
10 & 40 & $1.65 \pm 0.33$ & $0.55 \pm 0.21$ & $1.37 \pm 0.26$ & $0.52 \pm 0.33$ \\
20 & 60 & $1.96 \pm 0.33$ & $0.68 \pm 0.14$ & $1.29 \pm 0.26$ & $0.43 \pm 0.37$ \\
40 & 30 & $1.24 \pm 0.33$ & $0.27 \pm 0.24$ & $1.56 \pm 0.26$ & $0.68 \pm 0.23$ \\
40 & 40 & $1.66 \pm 0.33$ & $0.56 \pm 0.20$ & $1.53 \pm 0.26$ & $0.65 \pm 0.24$ \\
40 & 60 & $1.99 \pm 0.33$ & $0.70 \pm 0.13$ & $1.34 \pm 0.26$ & $0.49 \pm 0.34$ \\
60 & 40 & $1.39 \pm 0.33$ & $0.40 \pm 0.31$ & $1.57 \pm 0.26$ & $0.68 \pm 0.22$ \\
60 & 50 & $1.99 \pm 0.33$ & $0.70 \pm 0.13$ & $1.51 \pm 0.26$ & $0.64 \pm 0.25$ \\
60 & 60 & $2.05 \pm 0.33$ & $0.72 \pm 0.13$ & $1.24 \pm 0.26$ & $0.38 \pm 0.25$ \\
80 & 50 & $2.44 \pm 0.33$ & $0.82 \pm 0.09$ & $1.24 \pm 0.26$ & $0.38 \pm 0.25$ \\
90 & 50 & $2.14 \pm 0.33$ & $0.74 \pm 0.11$ & $1.84 \pm 0.26$ & $0.85 \pm 0.16$ \\
17.7 & 34.1 & $1.46 \pm 0.33$ & $0.44 \pm 0.28$ & $1.98 \pm 0.26$ & $0.91 \pm 0.13$ \\
42.5 & 49.0 & $1.60 \pm 0.33$ & $0.53 \pm 0.23$ & $1.38 \pm 0.26$ & $0.53 \pm 0.32$ \\
\enddata
\end{deluxetable}  

Table 2 shows a number of candidate directions for $\hat{b}$ with marginally significant evidence for anisotropy.  Directions such as $(\lambda=40^{\circ},\beta=40^{\circ})$, $(\lambda=40^{\circ},\beta=60^{\circ})$ (the case shown in Figure 2), $(\lambda=60^{\circ},\beta=50^{\circ})$, and $(\lambda=90^{\circ},\beta=50^{\circ})$ have values of $R(a=0.3)$ which exceed unity by 2$\sigma$ or more. A plot of $\xi^2$ vs $|\cos A|$ for $(\lambda=90^{\circ},\beta=50^{\circ})$ is shown in Figure 4, in the same format as Figure 2. 
\begin{figure}[h]
\epsscale{0.60}
\includegraphics[width=22pc,angle=-90]{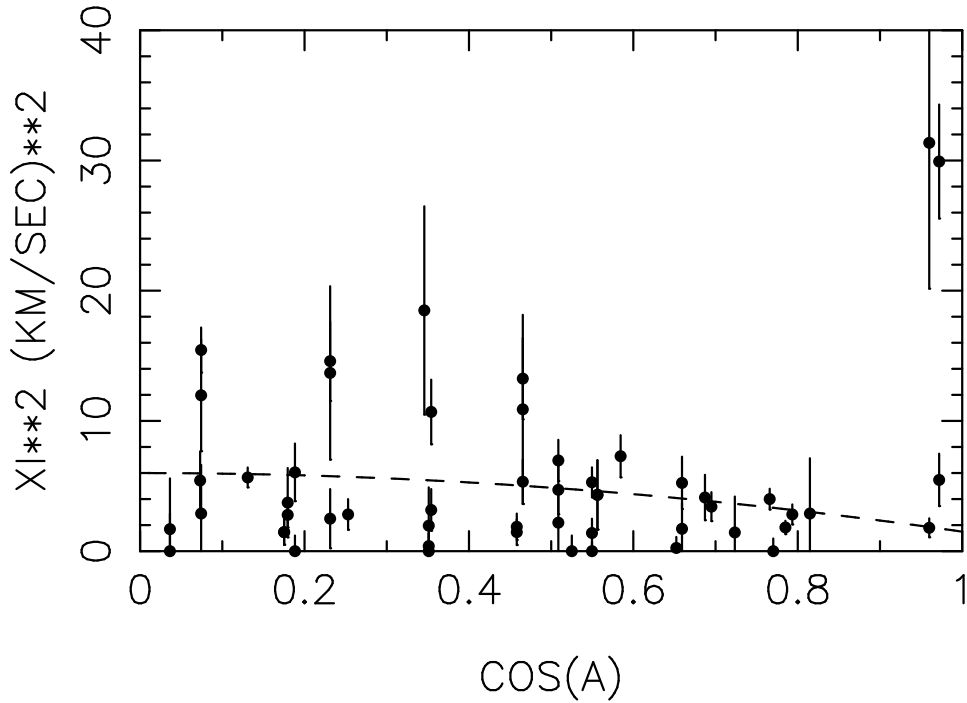}
\caption{Plot of $\xi^2$ versus $| \cos A |$ for $\lambda =90^{\circ}, \beta = 50^{\circ}$. The plot is in the same format as Figure 2. The dashed line shows the model given by Equation (3) with $V_{\perp}^2=6.0$ km$^2$/sec$^2$, and $\epsilon = 0.75$. }
\end{figure} 

In some of these cases both $R(a=0.3)$ and $R(a=0.5)$ exceed unity by about 2$\sigma$ or greater.  Furthermore, the inferred values for $\epsilon$ for the 2 binning intervals are in agreement, within the errors. Although the results of this analysis do not present strong evidence for anisotropy of the Local Cloud turbulence (our best cases are, after all, shown in Figures 2 and 4), they are not inconsistent with $\hat{b}$ pointing in the direction $(\lambda=40^{\circ} \pm 20^{\circ},\beta=50^{\circ} \pm 20^{\circ})$, and an anisotropy parameter $\epsilon = 0.5 - 0.7$. Before leaving this section, two points should be emphasized.  First, the modest indications of anisotropy in Table 2 are completely dependent on excluding the measurements of Alkaid and $\zeta$ Dor, which so prominently depart from the model curves in Figures 2 and 4.  Second, the measurements of $\xi^2$ vs. $|\cos A|$ do not adhere closely to the relationship given by Equation (3), but show a dispersion about that curve which is larger than the measurement error.  If anisotropy is present in these data, there must be another, unnamed physical process responsible for variation in $\xi$ from one line of sight to another.  
\subsubsection{The Value of $\epsilon$ for the Solar Wind}
Having presented our results on upper limits to the anisotropy factor for turbulence in the Local Clouds, we now consider the corresponding quantity in the solar wind, which has the benefit of direct, in-situ measurements.  \cite{Spangler04} resolved solar wind magnetic field fluctuations into components parallel and perpendicular to the large-scale interplanetary magnetic field. The data came from the magnetometer of the WIND spacecraft \footnote{WIND is one of the spacecraft which comprises the International Solar-Terrestrial Physics (ISTP) program.} at a heliocentric distance of about 1 AU.  \cite{Spangler04} used 66 intervals of one hour duration during slow solar wind conditions, and report their results in terms of modulation indices $m_{B \parallel}$ and $m_{B \perp}$ of fluctuations parallel and perpendicular to the mean field \footnote{\cite{Spangler04} used the variables $\epsilon_{B \parallel}$ and $\epsilon_{B \perp}$ for the modulation indices, but we do not retain this notation so as to avoid confusion with our anisotropy parameter $\epsilon$.}, 
\begin{eqnarray}
m_{B \parallel}= \frac{\delta b_{\parallel}}{B_0} \\
m_{B \perp} = \frac{\delta b_{\perp}}{B_0}
\end{eqnarray}
where $\delta b_{\parallel}$ and $\delta b_{\perp}$ are the rms values of the fluctuations in the magnetic field components parallel and perpendicular, respectively, to the mean field $B_0$.  \cite{Spangler04} report mean values for $m_{B \parallel}$ and $m_{B \perp}$ of 0.0321 and 0.112, respectively.  The means are for the distribution of values measured in the 66 data intervals.  As noted in \cite{Spangler04}, $m_{B \perp}$ should be larger than $m_{B \parallel}$ because it possesses contributions from two turbulent field components rather than just one.  The degree of anisotropy can be determined by comparing $m_{B \parallel}$ to $m_{B \perp}/\sqrt 2$.  \cite{Spangler04} found that the turbulent fluctuations in their study were Alfv\'{e}nic, in the sense that $\frac{\delta v}{V_A} = \frac{\delta b}{B_0}$, so the measured magnetic field modulation indices may be considered proxies for modulation indices of the velocity fluctuations.  With this assumption, we have 
\begin{eqnarray}
\frac{\sqrt 2 m_{B \parallel}}{m_{B \perp}} = \left( \frac{V_{\parallel}}{V_{\perp}} \right) \\
\epsilon = 1 - \left( \frac{V_{\parallel}}{V_{\perp}} \right)^2 = 1 - \frac{2 m_{B \parallel}^2}{m_{B \perp}^2}
\end{eqnarray}
where in Equations (8) and (9) we make the connection between the root-mean-square velocity fluctuations and the velocity scales $V_{\parallel}$ and $V_{\perp}$ of the fluctuation distribution function in the Appendix.  Using the values of  $m_{B \parallel}$ and $m_{B \perp}$ from \cite{Spangler04} in Equation (9), we have $\epsilon = 0.84$.  

The anisotropy of solar wind fluctuations had been considered prior to \cite{Spangler04} by \cite{Bavassano82} and \cite{Klein93}.  \cite{Bavassano82} also studied magnetic field fluctuations, in conditions of high speed solar wind at several heliocentric distances.  The results of \cite{Bavassano82} (see data in their Figure 2) yield values for $\epsilon$ in the range 0.8 - 0.9 and greater, i.e. very similar to that quoted above.  It should also be noted that a highly anisotropic and Alfv\'{e}nic nature is a characteristic of turbulence in the inner solar system, that might not be valid throughout interplanetary space.  \cite{Bavassano82} and \cite{Klein93} show that properties of solar wind turbulence depend on heliocentric distance, specifically that the degree of anisotropy decreases with increasing heliocentric distance.  \cite{Klein93} further report that the anisotropy is less in slow speed than high speed solar wind. The upper limit we can place to anisotropy of velocity fluctuations in the Local Clouds is less than, though comparable to, that of solar wind turbulence at a heliocentric distance of 1 AU.  
\subsection{Anisotropy in the Ion Temperature}
As discussed in Section 3, the analysis of line widths by \cite{Redfield04} also yields the ion temperature $T$.  Strictly speaking, this is a line-of-sight temperature; it is a measure of the line-of-sight component of thermal motion of atoms and ions.  If the Local Cloud turbulence is similar to heliospheric turbulence, the ion temperature $T$ might also depend on the angle $A$ between the line-of-sight and the local interstellar magnetic field, as a consequence of $T_{\perp} \neq T_{\parallel}$.  In regions of the solar wind in which ion cyclotron resonance heating appears to be active, and in particular in the solar corona, $T_{\perp} \gg T_{\parallel}$ \citep{Cranmer02}. \cite{Kasper09} have made an extensive study of the perpendicular-to-parallel temperature ratio $R_p = \frac{T_{\perp}}{T_{\parallel}}$ for protons in the solar wind.  Although it is not the case that $R_p > 1$ in all cases, it is true that $T_{\perp}$ is larger than it would be in an adiabatically-expanding solar wind.  Furthermore, \cite{Kasper09} identify boundaries in a $(R_p, \beta)$ plane, where $\beta$ is the conventional plasma physics quantity of the ratio of thermal to magnetic pressure. These boundaries are clearly seen in the distribution of measured values of $R_p$ and $\beta$, and correspond to instability thresholds for generation of plasma waves. Apparently these plasma waves heat the protons in a way which keeps them ``in bounds'' in a subset of the $(R_p, \beta)$ plane. The important point here is that in the corona and solar wind, the perpendicular and parallel temperatures are usually not equal, and $T_{\perp} > T_{\parallel}$ where ion cyclotron resonant effects are important.  

 To test for temperature anisotropy in the Local Clouds, an analysis similar to that of Section 4.1 was undertaken.  Plots of $T$ as a function of $\cos A$ were made for all 361 candidate directions for $\hat{b}$. These plots were visually examined for indicators that the data were organized according to a relation like that in Equation (3).  In the case of temperature anisotropy, the line-of-sight temperature should obey a relationship like Equation (3).  No case of a convincing temperature anisotropy was found.  Figure 5 shows the results for the direction $(\lambda = 10^{\circ}, \beta = 40^{\circ})$. 
\begin{figure}[h]
\epsscale{0.60}
\vspace{2.0cm}
\includegraphics[width=22pc,angle=-90]{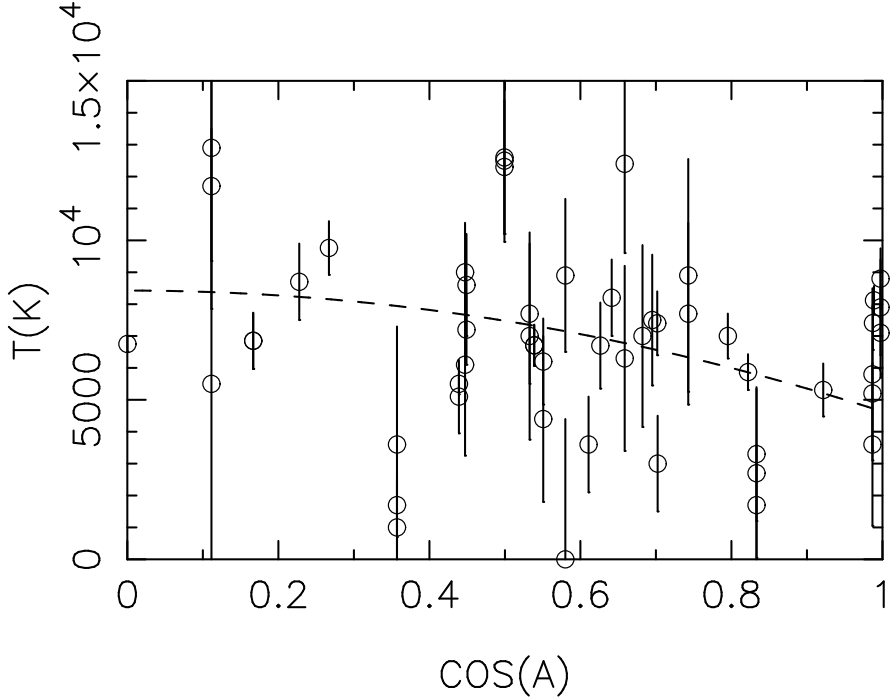}
\caption{The temperature as a function of $|\cos A|$ in the case of  $(\lambda = 10^{\circ}, \beta = 40^{\circ})$.  There is no obvious sign of temperature anisotropy, in the sense that $T_{\perp} > T_{\parallel}$. The dashed line represents a relationship similar to Equation (3), in which $T_{\perp}=8425K$, and $\epsilon=0.45$. }
\end{figure} 

An analysis similar to that of Section 4.1.1. was undertaken, in which a parameter $R_T$ was calculated for the mean temperature in two distinct intervals of width $a$.  Values for the ratio $R_T$ and associated errors were calculated for the same candidate directions as in Table 2.  The stars Alkaid and $\zeta$ Doradi were not excluded from this analysis because they are not anomalous as regards temperature. 
In none of these directions was there a convincing case for temperature anisotropy.  In only one of the cases examined, that of  $(\lambda = 10^{\circ}, \beta = 40^{\circ})$ shown in Figure 5, did both $R_T(a=0.3)$ and 
$R_T(a=0.5)$ exceed unity by an amount that was more than twice the adopted error.  For this direction, we calculate $R_T(a=0.3) = 1.45 \pm 0.08$ and $R_T(a=0.5) = 1.23 \pm 0.06$.  The errors were calculated from the dispersion in the measurements of $T$ given in \cite{Redfield04}, in a manner similar to that used in Section 4.1.1.  These values for $R_T$ would correspond to anisotropy parameters $\epsilon = 0.34 - 0.44$. We do not claim this direction as a case for temperature anisotropy because the data shown in Figure 5 do not show clear adherence to Equation (3).

For the other directions, the data indicate $R_T$ factors closer to unity, and anisotropy factors closer to zero.  We thus find no evidence in the data for an anisotropy in the sense $T_{\perp} > T_{\parallel}$, although (as illustrated by the discussion in the previous paragraph) we cannot exclude the possibility that $\epsilon \leq 0.40$ could be present, but hidden by random variations in $T$ from one line of sight to another.  An upper limit to the temperature anisotropy $\epsilon \leq 0.40$ corresponds to $\frac{T_{\perp}}{T_{\parallel}} \leq 1.67$, which is considerably less than that reported for the solar corona, and also less than many cases reported in the solar wind \citep[see Figure 1 of ][]{Kasper09}.
\subsection{Analysis of Data for Select Candidate Magnetic Field Directions}
The analysis of Section 4.1 was done in an unbiased fashion, i.e. with no a-priori estimate of the local direction of the interstellar magnetic field.  No direction examined had a compelling case for anisotropy of the turbulent amplitude $\xi$ or the ion temperature $T$.  With this analysis completed, we then re-examined the data for ``preferred'' candidate directions $\hat{b}$ advocated by \cite{Lallement05} and \cite{Opher09}, as discussed in Section 4.1 above.  

\cite{Lallement05} propose a direction of the local interstellar magnetic field of $205^{\circ} \leq \lambda \leq 240^{\circ}, -60^{\circ} \leq \beta \leq -38^{\circ}$.  Taking the means for each coordinates, we have a candidate field direction for \cite{Lallement05} of $(\lambda=222.5^{\circ},\beta=-49^{\circ})$.  Since $-\hat{b}$ serves equally well as a direction for anisotropy, we have $(\lambda=42.5^{\circ},\beta=49^{\circ})$ as a candidate direction for the local field.  It is interesting that this direction is very close to the set of directions chosen as the best candidates from the unbiased analysis of Section 4.1, and illustrated in Figure 2.  Although the weak, if not nonexistent, evidence of anisotropy in Figure 2 precludes any further claims, the coincidence of the set of directions chosen for closer examination and the proposed direction of the local field of \cite{Lallement05} could motivate future investigations with more lines of sight.  

The direction for $\hat{b}$ proposed by \cite{Opher09}, $(\lambda \simeq 15^{\circ},\beta \simeq 33^{\circ})$ did not emerge from our unbiased analysis of the $\xi$ data as one of the directions for closer examination. The values of $R(a=0.3)$ and $R(a=0.5)$ for a direction in the range of possible directions chosen by \cite{Opher09}, $(\lambda=17.7^{\circ},\beta=34.1^{\circ})$ are given in Table 2. The value of $R(a=0.3) = 1.46 \pm 0.33$ is not statistically significant, but is not inconsistent with an anisotropy of the velocity fluctuations $\leq 0.4 - 0.5$.  Interestingly, the anisotropy for $R(a=0.5)$ is larger, but not in agreement with the results for the smaller binning interval.  
\subsection{A Test for Ion Cyclotron Resonance Heating in the Local Clouds}
As noted in Section 2, in the solar corona and solar wind there is not a single temperature which is valid for all ions, as written in Equation (1).  Indeed, the temperature increases for ions with larger Larmor radii. The reason for adopting Equation (1) in application to the Local Clouds is the simple fact that it yields entirely satisfactory fits to the spectral line width data for lines from as many as 8 different atoms and ions (see Figure 1 of \cite{Redfield04}).  

The analysis of this section will be in the nature of establishing an upper limit to the ion mass dependence of the ion temperature in the Local Clouds.  If cyclotron resonant heating is occurring, one would expect a modification of Equation (1).  A plausible candidate form is 
\begin{equation}
b^2 = \frac{2 k_B T_0}{m} \left( \frac{m}{m_0}\right)^d + \xi^2
\end{equation}
where $T_0$ and $m_0$ are the temperature and mass of the lightest atom or ion analysed, and $m$ is the mass of the more massive atom or ion.  This equation essentially says that the atomic or ionic temperature $T(m) = T_0 (m/m_0)^d$. The form chosen for Equation (10) is relevant because, in the solar coronal case, the heating has been shown to be more pronounced than ``mass proportional heating'' \citep{Cranmer02}, which corresponds to $d \geq 1$. 

A fit of Equation (10) to the data introduces three model parameters ($T_0, \xi, d$), rather than the two parameters of Equation (1) utilized by \cite{Redfield04}.  This means that there is a broader basin of acceptability in a $\chi^2$ sense.  

The following analysis was undertaken. 
\begin{enumerate}
\item We selected data from all lines of sight and cloud components which possessed 7 or 8 transitions, including the deuterium (important for determining $T_0$) and iron (important for determining $\xi$) line measurements.  These data consisted of the measured line widths $b$ and associated errors. We had 11 such absorption components for analysis. 
\item A least-squares fit of Equation (10) was made to the data, and the range of parameters $T_0, \xi, \mbox{ and } d$ determined which allowed an acceptable value for the reduced $\chi_{\nu}^2$.  We chose a value of $\chi_{\nu}^2 = 2.21$, corresponding to a 5 \% probability of occurrence for 5 degrees of freedom \citep{Bevington69}, as the limit of an acceptable fit. 
\item Upper limits to $d$ were chosen which corresponded to the maximum value of $d$ consistent with the  $\chi_{\nu}^2 \leq 2.21$ acceptability criterion, with no imposed constraints on $T_0$ and $\xi$.  
\end{enumerate}  
\subsubsection{Results}
The results of this analysis are given in Table 3.  Column 1 gives the star name and column 2 gives the absorption component, identified by its velocity.  Columns 3 and 4 give the temperature and turbulent velocity parameter $\xi$ from the 2 parameter fit, Equation (1).  The numbers in these columns are taken directly from Table 1 of \cite{Redfield04}, and are reproduced here for comparison with the parameters of the model given by Equation (10).  The errors in columns 3 and 4 are approximations to those published by \cite{Redfield04}; those authors allowed for different errors above and below the mean value.  Columns 5,6, and 7 give the values of $T_0$, $\xi$, and $d$ for the limiting acceptable 3-parameter model as defined in point \# 3 of the previous section.  
\begin{deluxetable}{rrrrrrr}
\tabletypesize{\small}
\tablecaption{Limits on Mass Dependence of Ion Temperature}
\tablewidth{0pt}
\tablehead{\colhead{Star} & \colhead{Comp (km/sec)} & \colhead{$T$(K)} & \colhead{$\xi$(km/sec)} & \colhead{$T_0$(K)} & \colhead{$\xi$(km/sec)} & \colhead{$d$}}
\startdata
Capella & 21.5 & $6700 \pm 1400$ & $1.68 \pm 0.39$ & 6450 & 1.30 & 0.12 \\
$\iota$ Cap & -2.2 & $5500 \pm 5500$ & $3.70 \pm 0.90$ & 6020 & 2.82 & 0.57 \\
$\iota$ Cap & -12.1 & $12900 \pm 3800$ & $1.58 \pm 0.89$ & 5388 & 0.0 & 0.52 \\
$\iota$ Cap & -20.5 & $11700 \pm 4100$ & $3.82 \pm 0.44$ & 5490 & 0.0 & 0.74 \\
$\alpha$ Cent A & -18.4 & $5100 \pm 1200$ & $1.21 \pm 0.49$ & 3960 & 0.0 & 0.26 \\
G191-B2B & 8.6 & $4400 \pm 2800$ & $3.27 \pm 0.39$ & 3990 & 0.0 & 0.69 \\
G191-B2B & 19.2 & $6200 \pm 1400$ & $1.78 \pm 0.51$ & 5612 & 0.0 & 0.27 \\
HZ43 & -6.5 & $7500 \pm 2100$ & $1.70 \pm 1.70$ & 6235 & 0.0 & 0.32 \\
$\zeta$ Dor & 8.4 & $7700 \pm 2300$ & $2.34 \pm 0.48$ & 3826 & 0.0 & 0.61 \\
$\zeta$ Dor & 13.9 & $7000 \pm 3500$ & $5.47 \pm 0.41$ & 4296 & 0.41 & 1.00 \\
$\upsilon$ Peg & 1.7 & $1700 \pm 1100$ & $3.93 \pm 0.22$ & 1000 & 3.00 & 1.25 \\
\enddata
\end{deluxetable}  

Of the 11 absorption components with 7 or 8 measured transitions, 9 gave acceptable fits with the temperature model of Equation (10) and $d > 0$.  For two of the 11 components, the minimally-acceptable fit of Equation (10) yielded a physically implausible model for $b(m)$ in which $b$ increased with increasing $m$.  We will discuss those cases below.  

For the remaining 9 line of sight/absorption components, a minimally-acceptable fit to the data was possible with values of $d$ ranging from 0.12 (Capella) to 0.74 (-20.5 km/sec component of $\iota$ Cap).  It is worth emphasizing again that these values represent the maximum value of $d$ which is statistically acceptable as defined above, and with little or no constraints placed on the values of $T_0$ and $\xi$. Any larger value of $d$ is incompatible with the data. An illustration of a fit of Equation (10) to one of our data sets is shown in Figure 6. This figure also shows the best-fitting 2 parameter model, with the parameters reported by \cite{Redfield04}. We do not show the model with the maximally acceptable value of $d=0.69$, but instead Equation (10) with a slightly smaller value of $d=0.60$ (and associated parameters $T_0$ and $\xi$) that provides a better fit to the data.  
\begin{figure}[h]
\epsscale{0.60}
\includegraphics[width=20pc,angle=-90]{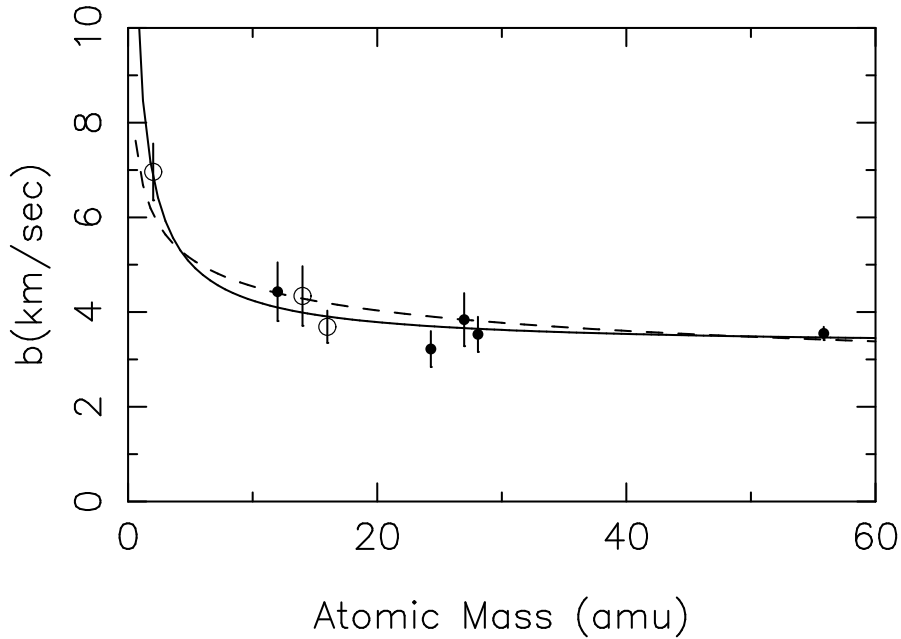}
\caption{Spectral line width data $b(m)$ for the 8.6 km/sec absorption component in the spectrum of G191-B2B.  Plotted is the measured spectral line width $b$ versus the atomic or ionic mass in atomic mass units. Solid symbols represent ions, open symbols are neutral atoms. The solid curve is the best fit of Equation (1) from \cite{Redfield04}; the parameters of the model are $T = 4400K$ and $\xi = 3.27$ km/sec. The dashed curve represents Equation (10) with $T = 4133K$, $\xi = 1.63$ km/sec, and $d = 0.60$. The reduced chisquared $\chi^2_{\nu}$ for the solid curve ($d=0$) is 0.64.  The dashed curve, corresponding to $d=0.60$, has $\chi^2_{\nu} = 1.82$, which is larger, but still statistically acceptable by our criteria.  }  
\end{figure}
In all 9 cases, the fit of Equation (10) introduced lower values of $\xi$ than the two-parameter fits of Equation (1) published in \cite{Redfield04}.  For example in the case of the 8.6 km/sec component of G191-B2B, \cite{Redfield04} report $\xi = 3.27$ km/sec, while the limiting acceptable fit of Equation (10) has $\xi = 0$ km/sec and $d = 0.69$. In fact, as may be seen in Table 3, most of the limiting acceptable fits of Equation (10) to the data have $\xi = 0$.   

The reason for this anticorrelation of $\xi$ and $d$ is clear.  The quantity $\xi$ is determined by the degree to which $b(m)$ asymptotes to a constant level as $m$ becomes large.  However, this behavior can also be produced (over a limited range in $m$) by a lower value of $\xi$, if compensated by a higher thermal width for more massive ions (the meaning of the $(m/m_0)^d$ term). 

The conclusion from this analysis is that the data do not exclude a modest dependence of the ion temperature on mass, with an associated drop in the turbulence level $\xi$.  This result is somewhat ironic in that ion-mass-dependent temperature was sought as a consequence of turbulence or wave-particle interaction.  However, if mass dependence of temperature is present, the data require a reduced level of turbulence in the Local Clouds.  

Two points should be emphasized in this context.  First, there is no statistical argument for choosing Equation (10) over the simpler Equation (1) as a temperature model; both provide statistically-acceptable representations of the data, and in almost all cases the best fit has $d=0$.  Second, with the exception of the two lines of sight with poor fits, the degree of mass-dependence of the temperature which is compatible with the data is less than the case of mass-proportional temperature ($d$ = 1) that is the benchmark for the corona.  

To conclude this section, we briefly discuss the two cases in which a plausible model for $T(m)$ according to Equation (10) was not obtained.  These cases are the 14 km/sec absorption component for $\zeta$ Dor, and the 1.7 km/sec component of $\upsilon$ Peg.  In the case of $\zeta$ Dor, the value of $\xi$ from \cite{Redfield04} was particularly high (5.47 km/sec, noted above) and in the case of $\upsilon$ Peg, the temperature was low (1700K, \cite{Redfield04}).  In both cases, therefore the measured variation of $b$ with $m$ was relatively small, so fit trends with $\frac{db}{dm} > 0$ or $\frac{db}{dm} < 0$ were equally compatible with the data. 
   
\subsection{Additional Remarks on Heating by Plasma Waves}
The previous section dealt with a search for mechanisms that produce more effective heating of ions with lower cyclotron frequencies.  Two mechanisms have been proposed which can explain this, both of which involve the interaction of ions with plasma waves and turbulence.  These are the ion-cyclotron resonance interaction discussed by \cite{Cranmer02} and \cite{Hollweg08} {\em inter aliis}, and the stochastic acceleration mechanism of \cite{Chandran10}. However, our results on spectral line widths as a function of ionic or atomic mass place more general constraints on plasma physics processes operative in the Local Clouds.  Such processes, like magnetic reconnection, interaction with plasma waves, response to large-scale electric fields, etc, act directly only on ions since ions respond to electric and magnetic fields, while neutral atoms do not.  The data of \cite{Redfield04} include both neutral atoms and ions. 
The data in Figure 6 include measurements for lines of CII, MgII, AlII, SiII, and FeII, which are ions (solid symbols), as well as lines of the neutral atoms DI, NI, and OI (open symbols).  In Figure 6, the two-parameter curve fits all transitions with the same temperature.  The more extensive set of results in Figure 1 of \cite{Redfield04} yields the same conclusion; a single temperature provides a satisfactory description of the data for both ions and neutral atoms. This result is obtained for many lines of sight throughout the sky.  This indicates that whatever process heats the Local Cloud plasmas to temperatures of order 7000K does so via a process which heats ions and neutrals equally.  Alternatively, if heating power is input preferentially or exclusively to ions via some plasma physics process, that power is efficiently shared with neutrals.

\section{Discussion}
In this section, we consider the observational findings of Section 4, and speculate on their consequences for our understanding of the turbulence in the Local Clouds. 
\subsection{Anisotropy of Velocity Fluctuations} Our analysis shows no compelling evidence that the turbulent velocity width $\xi$ depends on direction in the sky, in a way which would be consistent with velocity fluctuations predominantly perpendicular to a large scale magnetic field (see cartoon in Figure 1). The fact that the few directions chosen for closer inspection have $\hat{b}$ in approximately the same direction as that independently inferred by \cite{Lallement05} is an interesting curiosity and probably motivates future examinations with more data, but does not change the negative nature of the current conclusion.  We feel that the isotropy of $\xi$ is significant, and contains information on the nature of turbulence in the Local Clouds.  There are three possible explanations for this result, which are not mutally exclusive.  
\begin{itemize}
\item The simplest explanation is that turbulence in the Local Clouds is not Alfv\'{e}nic, so that the velocity fluctuations are not primarily perpendicular to the large scale field in the solar neighborhood.  The velocity fluctuations parallel to the large scale field are then comparable to those in the two perpendicular directions. The mechanisms responsible for decay of turbulent anisotropy in the remote solar wind (cited in Sections 2 and 4.1.2) should be studied as being potentially relevant to the Local Clouds.  
\item A related suggestion is that the fluctuations are transverse on scales much smaller than the outer scale \citep{Spangler99}, but that the measurements of $\xi$ are dominated by the largest amplitude fluctuations on the outer scale.  These outer scale fluctuations are of sufficient amplitude to nearly totally randomize the average field, and thus eliminate the simple observational signature sought for in Figure 1. 
\item The third explanation is closely related to the second, and would say that turbulent fluctuations nearly completely randomize the local Galactic magnetic on a scale comparable to the separation between the Local Clouds.  According to this viewpoint, Figure 1 would refer only to the average magnetic field, and not to the true value at a specific cloud (or even within a cloud).  In terms of the mathematical vocabulary introduced in Section 4.1, the unit vector $\hat{b}$ would be randomized on scales comparable to the spacing between clouds, if not within individual clouds.
\end{itemize} 
 The last two explanations would seem to be the most plausible and consistent with our understanding of turbulence.  It is, after all, the dominant interpretation of why Galactic Faraday rotation measurements give such an ambiguous indication of a large scale Galactic magnetic field \citep[e.g.][]{Rand89,Rand94,Beck96,VanEck11}.  
However, in the present context, there are arguments against their applicability.  

Acceptance of point \# 3 would require the outer scale of the interstellar turbulence (at least in the vicinity of the Sun) to be of the order of a few parsecs.  \cite{Minter96} estimate the outer scale of the {\em three dimensional} turbulence to be about 4 parsecs, but suggest that there is a two dimensional component with an outer scale of order 100 parsecs. Since the rms amplitude of the magnetic field fluctuations is dominated by the two dimensional component, the \cite{Minter96} estimate of the turbulent Galactic field suggests that the correlation length in $\hat{b}$ is substantially larger than the distances between the stars used in this study.  A rigorous analysis has not been done however.  

Estimates of the outer scale in interstellar turbulence which may lead to a different conclusion have been made by Haverkorn and her co-workers \citep[e.g.][]{Haverkorn04a,Haverkorn04b,Haverkorn06}; a summary of Haverkorn's results on the outer scale is given in \cite{Spangler10}. Haverkorn has presented results using both Faraday rotation of background radio sources as well as the statistics of polarization fluctuations in the Galactic polarized synchrotron radiation.  \cite{Haverkorn06} claim evidence for a difference in the properties of fluctuations in the spiral arms of the Galaxy relative to the interarm regions.  For the case of the fluctuations in the spiral arms, their estimates of the outer scale range from 2-17 parsecs.  There was no effort in this work to distinguish between 2D and 3D turbulence, with different outer scales.  As noted in \cite{Spangler10}, an outer scale as small as 2-17 parsecs (particularly the low end of that range) would probably randomize the local magnetic field sufficiently to eliminate anisotropy of $\xi$ on the sky. These considerations provide another reason for improved determination of the turbulent outer scale in the interstellar medium.  It should be kept in mind that the observational results of \citep{Haverkorn04a,Haverkorn04b,Haverkorn06} and \cite{Minter96} most probably refer to the Warm Ionized Medium (WIM) component of the interstellar medium, and that their results are not obviously transferable to the turbulence in the Local Clouds.  

A second argument against points \#2 and \#3 is that we have information on the dimensionless amplitude of the Local Cloud turbulence, and the amplitude is small.  The mean value of $\xi$ from the data in \cite{Redfield04} is 2.24 km/sec. The amplitude of turbulence has been discussed in Section 4.1.2 above, and for fluctuations of all kinds (parallel and perpendicular to a large scale field) can be defined as $m \equiv \delta v/V_A$.  We can use the data in Table 1 to calculate the Alfv\'{e}n speed in the Local Clouds.  The value obtained depends on whether the density used is the total density (ion plus neutral) or only the ionized component.  The appropriate value depends on the scale of the fluctuations involved.  For large-scale fluctuations (larger than the ion-neutral collisional mean free path) ion-neutral collisions cause the neutral atoms to be carried along in the Alfv\'{e}n wave, and the total density should be used.  For small scale fluctuations (scales much smaller than the ion-neutral mean free path) the Alfv\'{e}nic waves or fluctuations occur only in the ionized fluid, and the ionized density should be used.  
Since $\xi$ is an rms value formed from all fluctuations, the dominant contribution to $\xi$ should come from large scales of order the outer scale, and the relevant density for calculating the Alfv\'{e}n speed should be the total density.  We will use the total density in the calculations below. 

Given the above considerations and the data in Table 1, the Alfv\'{e}n speed in the Local Clouds is 12.8 - 17.0 km/sec, with the range reflecting the possible values for the magnetic field strength.  In calculating the dimensionless amplitude of the turbulence, $ \delta v/V_A$, we assume $\delta v = \sqrt 3 \xi$, since $\xi$ corresponds to only one component of the turbulent velocity fluctuations. For $\xi$ we use the mean of the entire sample of \cite{Redfield04}, $\bar{\xi} = 2.24$ km/sec, giving $\delta v = 3.88$ km/sec. We then have for the dimensionless amplitude of the turbulence $0.23 \leq  \delta v/V_A \leq 0.30$. Once again, for Alfv\'{e}nic turbulence,  $\delta b/B_0$ would have the same range.  

These calculations indicate that the turbulence in the Local Clouds is of substantial amplitude, but probably not so large as to cause significant excursions of the magnetic field direction, i.e. variations in field direction $\leq 17^{\circ}$. These calculations argue that turbulence in these clouds is not sufficiently large to cause randomization of the field direction and thereby produce loss of intrinsic velocity anisotropy, if it were present.

\subsection{Limits on Ion Cyclotron Resonance Heating}
Perhaps the most striking difference between coronal and solar wind turbulence on one hand, and that in the Local Clouds on the other, is the absence in the Local Clouds of collisionless plasma processes responsible for enhanced heating of ions with larger Larmor radii (see Figure 6).  The existence of a single temperature for many ions (and neutral atoms) is an observational result which can not be an ``artifact'' of a randomized interstellar magnetic field.  In this section, we conjecture on the physical processes responsible for this apparent thermal equilibrium. A preliminary discussion of this topic has been given by \cite{Spangler11}.  

The simplest way of explaining this result is to invoke collisionality.  The solar corona is highly collisionless, and the solar wind at 1 au ranges from collisionless to only weakly collisional.  Although the density of the Local Cloud plasmas is lower than either of the heliospheric plasmas we have discussed, the temperature is also lower, leading to higher collision frequencies. More importantly, the Local Cloud plasmas possess a significant neutral component (see Table 1), so ion-neutral collisions occur. The role of collisions is to take energy input by collisionless processes into one ion species, and into a limited number of degrees of freedom (i.e. perpendicular, but not parallel motions), and redistribute it to many species and all degrees of freedom.  Observational support for the idea that collisions will fulfill this role is given by solar wind observations reported by \cite{Kasper09}.  More collisional parts of the solar wind, such as the heliospheric current sheet, lack properties such as mass-proportional temperature and temperature anisotropy.   We now consider whether the ion-neutral collision frequency in the Local Clouds is sufficiently large to make these plasmas collisional. 

\cite{Spangler11} use two definitions of collisionality employed by \cite{Uzdensky07} in the context of magnetic field reconnection in the solar corona.  
\begin{enumerate}
\item According to the first criterion, a collisionless plasma is one for which the ion cyclotron frequency is higher than the ion collision frequency.  This permits, for example, instabilities with growth rates of the order of a fraction of the cyclotron frequency to develop without modification by collisions.  
\item In the second, much more restrictive criterion, a collisionless plasma is one for which the collisional mean free path is larger than the dimensions or characteristic scale of the plasma or medium.  The converse situation of a collisional mean free path much smaller than the size of the system would then constitute a collisional plasma, even if the medium satisfied the first criterion for a collisionless plasma given above. 
\end{enumerate}

To evaluate the collisionality of the Local Cloud plasmas, we need to identify the relevant collisional processes.  \cite{Spangler11} discuss charge exchange and induced dipole scattering.  These two processes have similar collision cross sections, with charge exchange being larger by a factor of a few for Local Cloud conditions.  The microphysics of what occurs is quite different in the two cases.  For conditions appropriate to the Local Clouds, \cite{Spangler11} calculate an H$^+$-H ion-neutral collision frequency due to induced dipole scattering of $3 \times 10^{-10}$ Hz.  This frequency is about 8 orders of magnitude smaller than the proton ion gyrofrequency \citep[see Table 2 of ][]{Spangler11}, meaning that the Local Clouds easily satisfy the first of the above criteria for being collisionless.  However, the collisional mean free path corresponding to this collision frequency is $5 \times 10^{15}$ cm = $1.5 \times 10^{-3}$ parsec = 330 AU.  For the case of H$^+$-H charge exchange, the collision frequency would be a few times higher, and the collisional mean free path a few times smaller than the numbers given immediately above.  In any case, we conclude that the Local Clouds are highly collisional by Uzdensky's second criterion.  

It seems likely that collisions between ions taking part in the Alfv\'{e}nic waves and turbulence in the local clouds, and neutral atoms which are not, is responsible for the removal of temperature anisotropy and ion-specific heating. It would be worthwhile to conduct a theoretical study of ion and neutral atom dynamics in the presence of Alfv\'{e}nic turbulence, including collisions due to both charge exchange and induced dipole scattering. Such calculations could determine if the collisions do indeed remove temperature anisotropy and ion-specific temperatures, by taking the energy which is preferentially input to one or a few ions, and distributing it to all ion and neutral atom species.  

Introducing Uzdensky's first criterion of collisionality has provided interesting insight in the present context. The fact that the collision frequency is approximately 8 orders of magnitude smaller than the gyrofrequency means that plasma instabilities and quasilinear modification of ion distribution functions would have ample time to develop, unencumbered by collisions, if conditions for instabilities were present.  The absence of ``collisionless'' observational signatures suggests that energetic processes on ion temporal and spatial scales are not occurring in the Local Clouds.     
 
\section{Summary and Conclusions}
The conclusions of this paper are as follows.
\begin{enumerate}
\item High-resolution optical and UV spectroscopy of absorption lines for several ions and atoms in the Local Clouds of the VLISM provide a remarkable number of diagnostics of the turbulence in these clouds.  The data are sufficient to make some tests of the similarities to, or differences from, MHD turbulence in the solar corona and solar wind. 
\item The turbulent line broadening parameter $\xi$ does not show a systematic dependence on direction in the sky, as would be expected if the velocity fluctuations are predominantly transverse to the local Galactic magnetic field.  This result indicates either that the turbulent fluctuations are not anisotropic with respect to the local magnetic field, or that the field is partially randomized on a size scale of several parsecs. 
\item The quantitative limits we can place on anisotropy in the case of an ordered field are, however, not very restrictive. Our results indicate an upper limit to the anisotropy parameter $\epsilon \leq 0.70$.  If such a large value of the anisotropy is present, it must be masked by another, unidentified process that causes variations in $\xi$ from one line of sight to another, which in many cases are within the same cloud.
\item Although a conclusion of this paper is that there is no significant anisotropy in $\xi$, and therefore no result for the direction of the local interstellar magnetic field $\vec{B}_0$, a set of candidate directions for $\vec{B}_0$ chosen for further analysis are in rough agreement with the field direction proposed by \cite{Lallement05}.  This could justify future investigation with a larger set of absorption line measurements.  
\item The temperature $T$ obtained from the spectral line widths (see Equation (1)) also shows no dependence on direction on the sky. The most straightforward interpretation of this result is that there is not a temperature anisotropy $T_{\perp} > T_{\parallel}$ as is the case in the solar corona, and to a lesser extent, the solar wind. 
\item We determined an upper limit to the amount of ion-mass-dependent heating by studying the statistical acceptability of Equation (10), a temperature model with mass-dependent temperature.  This model was fit to the data for 9 lines of sight with 7 or 8 transitions from ions or atoms of different mass.  The maximum value of $d$ allowed for these 9 lines of sight was $d=0.74$, with the other lines of sight having smaller values of $d$.  The case of mass-proportional temperature would have $d=1$.  
\item The explanation for results \# 3-5 is most likely that the turbulence (and ion and neutral atom distribution functions) have persisted for many ion-neutral collision timescales. These ion-neutral collisions have apparently eliminated temperature anisotropies and temperature differences between species.  This conclusion itself is of interest in that it indicates the absence of energetic kinetic plasma physics processes on the scale of the ion cyclotron radius or ion inertial length in the Local Clouds of the Very Local Interstellar Medium.   
\end{enumerate}

\acknowledgements
This work was supported at the University of Iowa by grants AST09-07911 and ATM09-56901 from the National Science Foundation. 
\appendix 
\section{A model for turbulent fluctuations}
In this appendix, we derive the relationship between the probability distribution function of velocity fluctuations in a frame of reference defined by the interstellar magnetic field, and the distribution function in an observer-centered frame defined by the line-of-sight to a star.  

The anisotropic turbulent velocity fluctuations are most naturally defined in a coordinate system in which one coordinate, the $z$ coordinate, is in the direction of the large scale magnetic field $\vec{B}_0$.  Let $\hat{b}$ be the unit vector which points in the direction of $\vec{B}_0$. The unit vector $\hat{b}$ can be defined by the galactic longitude and latitude $(\lambda,  \beta)$ towards which it points.  In the $\vec{B}_0$ coordinate system, the $z$ axis is in the direction of $\hat{b}$, $x$ is in the plane defined by $\hat{b}$ and the direction to the north galactic pole, and $y$ completes a right-handed coordinate system.  

The turbulence model we want to test is one in which the fluctuations in the plane perpendicular to $\hat{b}$ are larger than those in the direction of $\hat{b}$. A probability distribution function which is simple in mathematical form and describes this is 
\begin{equation}
f(\vec{v}) = \left( \frac{1}{(2 \pi)^{3/2} V_{\parallel} V_{\perp}^2} \right) \exp (-\frac{v_z^2}{2 V_{\parallel}^2}) \exp (-\frac{v_x^2 + v_y^2}{2 V_{\perp}^2}) 
\end{equation}
with $V_{\perp} > V_{\parallel}$ by assumption.  The distribution function (A1) satisfies the normalization requirement  $\int d^3 v f(\vec{v}) = 1$. 

The observed spectral line width is proportional to the rms fluctuation in the component of the velocity along the line of sight, which is in the direction of the unit vector $\hat{l}$.  The unit vector $\hat{l}$ points in the direction of galactic longitude and latitude $(l,b)$.  In what follows, we assume that the observed turbulent width $\xi$ can be expressed as
\begin{equation}
\xi^2 = <v_L^2> = <v_L^2(v_x,v_y,v_z)> = \int d^3v v_L^2 (v_x,v_y,v_z) f(\vec{v})
\end{equation}
where $v_L$ is the line-of-sight component of the flow velocity.  
By expressing $\xi^2$ as an expectation value, we assume that the line of sight integration through the cloud samples a large number of independent eddies in the cloud and thus satisfies the ergodic theorem.  

To evaluate the expression (A2), we need to express $v_L(v_x,v_y,v_z)$.  We define a coordinate system such that the unit vector $\hat{e}_L$ for one coordinate is in the direction of $\hat{l}$.  The unit vector corresponding to another coordinate ($\hat{e}_T$) is in the plane defined by $\hat{e}_L$ and the direction to the north galactic pole. Finally, the unit vector for the third coordinate in the line-of-sight coordinate system, $\hat{e}_P$, is defined by $\hat{e}_T \times \hat{e}_P = \hat{e}_L$.  

The transformation between the two coordinate systems is given by 
\begin{equation}
\left( \begin{array}{c}
v_T \\
v_P \\
v_L
\end{array} \right) = T \cdot \left( \begin{array}{c}
v_x \\
v_y \\
v_z
\end{array} \right)
\end{equation} 
where $T$ is the matrix which generates the rotation from the magnetic-field oriented coordinate system to the line-of-sight coordinate system.  The matrix $T$ is given by an Euler angle transformation, so it is the product of three matrices, each describing the rotation through one of the Euler angles, 
\begin{equation}
T = T_3 \otimes T_2 \otimes T_1
\end{equation}
The $T_1$ operator is the first one.  It rotates the $(x,y,z)$ coordinate system about the $y$ axis so that the $z^{'}$ axis is in the galactic plane. It is given by 
\begin{equation}
T_1 = \left( \begin{array}{ccc}
\cos \beta &  0 & \sin \beta  \\
0  &  1  &  0  \\
-\sin \beta &  0  & \cos \beta 
\end{array}  \right)
\end{equation}
The next rotation operator $T_2$ rotates the coordinate system in the galactic plane such that the $z^{''}$ axis is pointing along the galactic longitude $l$ of the line of sight.  It effects a rotation about the $x^{'}$ axis through an angle $\Delta l \equiv \lambda - l$ 
\begin{equation}
T_2 = \left( \begin{array}{ccc}
 1 & 0 & 0  \\
 0 & \cos \Delta l & -\sin \Delta l  \\
 0 & \sin \Delta l & \cos \Delta l 
\end{array}  \right)
\end{equation}
The third rotation operator rotates the coordinate system ``upwards'' about the $y^{''}$ axis through an angle $b$.  At this point, the $z^{'''}$ direction coincides with $\hat{l}$.  The operator $T_3$ is given by
\begin{equation}
T_3 = \left( \begin{array}{ccc} 
 \cos b & 0 & -\sin b  \\
 0 & 1 & 0  \\
 \sin b & 0 & \cos b 
\end{array}  \right)
\end{equation}
Multiplying the three matrices together as in Equation (A4) to produce $T$, we have
\begin{equation}
T = \left( \begin{array}{ccc}
 \cos b \cos \beta + \cos \Delta l \sin b \sin \beta & -\sin \Delta l \sin b & \cos b \sin \beta - \cos \Delta l \sin b \cos \beta \\
 \sin \Delta l \sin \beta & \cos \Delta l & -\sin \Delta l \cos \beta   \\
\sin b \cos \beta - \cos \Delta l \cos b \sin \beta & \sin \Delta l \cos b & \sin b \sin \beta + \cos \Delta l \cos b \cos \beta  
\end{array}  \right)
\end{equation}
We are interested in only the $v_L$ component in the (T,P,L) coordinate system, which is given by 
\begin{equation}
v_L = (\sin b \cos \beta -\cos \Delta l \cos b \sin \beta)v_x + (\sin \Delta l \cos b)v_y + (\sin b \sin \beta + \cos \Delta l \cos b \cos \beta)v_z
\end{equation}
In keeping with our model of the clouds and their turbulence, Equation (A9) holds at every point along the line of sight through a cloud. 
The quantity which is measured as the turbulence parameter, $\xi^2 = <v_L^2>$.  We assume that $<v_x v_y> = <v_x v_z> = <v_y v_z> = 0 $ (i.e. the fluctuations have no dominant polarization).  We then have 
\begin{eqnarray}
\xi^2 = <v_L^2> = (\sin b \cos \beta -\cos \Delta l \cos b \sin \beta)^2 <v_x^2> + (\sin \Delta l \cos b)^2 <v_y^2> \\ \nonumber 
+ (\sin b \sin \beta + \cos \Delta l \cos b \cos \beta)^2 <v_z^2>
\end{eqnarray}
The expectation values of the squares of the velocity components are
\begin{equation}
<v_i^2> \equiv \int d^3v v_i^2 f(\vec{v})
\end{equation} 
$i = x, y, z$, and for our model $<v_x^2> = <v_y^2>$.  Substituting Equation (A1) into (A11) and evaluating gives 
\begin{eqnarray}
<v_x^2> = <v_x^2> = V_{\perp}^2 \\
<v_z^2> = V_{\parallel}^2
\end{eqnarray}
We now define a first anisotropy parameter $\eta \equiv \frac{V_{\parallel}^2}{V_{\perp}^2}$, and use it in Equation (A10) to give  
\begin{equation}
\frac{\xi^2}{V_{\perp}^2} = \frac{<v_L^2>}{V_{\perp}^2} = \left[ (\sin b \cos \beta - \cos \Delta l \cos b \sin \beta)^2 + (\sin \Delta l \cos b)^2 + \eta (\sin b \sin \beta + \cos \Delta l \cos b \cos \beta)^2 \right]
\end{equation}
This is an appealing equation.  The measured square of the turbulent velocity width is on the left hand side of the equation, and functions of galactic coordinates (of the line of sight and the large scale B field), as well as the anisotropy parameter $\eta$ are on the right.  However, the expression can be simplified further by use of trigonometric identities, and definition of a second anisotropy parameter $\epsilon$, $\eta \equiv 1 - \epsilon$, to be
\begin{equation}
\frac{\xi^2}{V_{\perp}^2} = \frac{<v_L^2>}{V_{\perp}^2} = 1 - \epsilon (\sin b \sin \beta + \cos \Delta l \cos b \cos \beta )^2
\end{equation}
This is one of the two fundamental equations in our analysis.  Equation (A15) relates the observed turbulent line width $\xi$ (left hand side, normalized by the unknown but estimatable $V_{\perp}^2$) to the anisotropy parameter $\epsilon$, the direction of the line of sight $(l,b)$ and the direction of the local interstellar magnetic field $(\lambda, \beta)$ on the right hand side. Although this form of the equation is most useful for the analyses carried out in this paper, it is more instructive to rewrite Equation (A15) in terms of an angle $A$, defined by $\hat{l} \cdot \hat{b} = \cos A$.  It is easy to show that Equation (A15) becomes 
\begin{equation}
\frac{\xi^2}{V_{\perp}^2} = \frac{<v_L^2>}{V_{\perp}^2} = 1 - \epsilon \cos^2A
\end{equation}
One of the goals of this paper is to determine if, for some set of values of $(\lambda, \beta)$, Equation (A15) provides a good representation of the measured values of $\xi$ for some value of $\epsilon \geq 0$.


\begin{thebibliography}{}
\bibitem[Bavassano et al (1982)]{Bavassano82} Bavassano, B., Dobrowolny, M., Fanfoni, G., Mariani, F., and Ness, N.F. 1982, \solphys~78, 373  
\bibitem[Bayley et al (1992)]{Bayley92} Bayley, B.J., Levermore, C.D., and Passot, T. 1992, Phys. Fl. A~4, 945
\bibitem[Beck et al (1996)]{Beck96} Beck, R., Brandenburg, A., Moss, D., Shukurov, A., and Sokoloff, D. 1996, \araa~34, 155  
\bibitem[Bevington (1969)]{Bevington69} Bevington, P.R. 1969, {\em Data Reduction and Error Analysis for the Physical Sciences}, (McGraw-Hill: New York), p313 
\bibitem[Bird and Edenhofer (1990)]{Bird90} Bird, M.K. and Edenhofer, P. 1990, in {\em Physics of the Inner Heliosphere}, R. Schwenn and E. Marsch (ed), (Springer-Verlag:Berlin), p13
\bibitem[Bruno and Carbone (2005)]{Bruno05} Bruno, R. and and Carbone, V. 2005, Living Reviews Solar Phys.~2,4
\bibitem[Chandran (2010)]{Chandran10} Chandran, B.D.G. 2010, \apj~720, 548
\bibitem[Cranmer (2002)]{Cranmer02} Cranmer, S.R. 2002, \ssr~101, 229 
\bibitem[Frisch (2000)]{Frisch00} Frisch, P.C. 2000, Am. Sci.~88, 52 
\bibitem[Frisch et al (2011)]{Frisch11} Frisch, P.C., Redfield, S., and Slavin, J.D. 2011, \araa (in press) 
\bibitem[Goldstein et al (1995)]{Goldstein95} Goldstein, M.L., Robert, D.A., and Matthaeus, W.H. 1995, \araa~33, 283
\bibitem[Gurnett et al (2006)]{Gurnett06} Gurnett, D.A., Kurth, W.S., Cairns, I.H., and Mitchell, J. 2006, in {\em Physics of the Inner Heliosheath}, American Institute of Physics Conference Proceedings \# 858, p129 
\bibitem[Haverkorn et al (2004a)]{Haverkorn04a} Haverkorn, M., Katgert, P., and de Bruyn, A.G. 2004a, \aap~427, 169
\bibitem[Haverkorn et al (2004b)]{Haverkorn04b} Haverkorn, M., Gaensler, B.M., McClure-Griffiths, N.M., Dickey, J.M., and Green, A.J. 2004b, \apj~609, 776 
\bibitem[Haverkorn et al (2006)]{Haverkorn06} Haverkorn, M., Gaensler, B.M., Brown, J.C., Bizunok, N.S., McClure-Griffiths, N.M., Dickey, J.M., and Green, A.J. 2006, \apj~637, L33  
\bibitem[Hollweg (2008)]{Hollweg08} Hollweg, J.V. 2008, J. Astrophys. Astr.~29, 217
\bibitem[Kasper et al  (2009)]{Kasper09} Kasper, J.C., Maruca, B.A., and Bale, S.D. 2009, arXiv:0911.2715
\bibitem[Klein et al  (1993)]{Klein93} Klein, L., Bruno, R., Bavassano, B., and Rosenbauer, H. 1993, \jgr~98, 17461
\bibitem[Lallement et al (2005)]{Lallement05} Lallement, R., Quemerais, E., Bertaux, J.L., Ferron, S., Koutroumpa, D., and Pellinen, R. 2005, Science~307, 1447
\bibitem[Minter and Spangler (1996)]{Minter96} Minter, A.H. and Spangler, S.R. 1996, \apj~458, 194
\bibitem[Minter and Spangler (1997)]{Minter97} Minter, A.H. and Spangler, S.R. 1997, \apj~485, 182
\bibitem[Moebius (2009)]{Moebius09} Moebius, E. et al 2009, \ssr~146, 149
\bibitem[Moran (2003)]{Moran03} Moran, T.G. 2003, \apj~598, 657
\bibitem[Opher et al (2009)]{Opher09} Opher, M. et al  2009, \nat~462, 1036
\bibitem[Rand and Kulkarni (1989)]{Rand89} Rand, R.J. and Kulkarni, S.R. 1989, \apj~343, 760 
\bibitem[Rand and Lyne (1994)]{Rand94} Rand, R.J. and Lyne, A.G. 1994, \mnras~268, 497
\bibitem[Redfield and Linsky (2001)]{Redfield01} Redfield, S. and Linsky, J.L. 2001, \apj~551, 413 
\bibitem[Redfield and Linsky (2004)]{Redfield04} Redfield, S. and Linsky, J.L. 2004, \apj~613, 1004
\bibitem[Redfield and Linsky (2008)]{Redfield08a} Redfield, S. and Linsky, J.L. 2008, \apj~673, 283
\bibitem[Redfield and Falcon (2008)]{Redfield08b} Redfield, S. and Falcon, R.E. 2008, \apj~683, 207
\bibitem[Redfield (2009)]{Redfield09} Redfield, S. 2009, \ssr~143, 323
\bibitem[Scudder (1992)]{Scudder92} Scudder, J.D. 1992, \apj~398, 299
\bibitem[Spangler (1999)]{Spangler99} Spangler, S.R. 1999, \apj~522, 879
\bibitem[Spangler and Spitler (2004)]{Spangler04} Spangler, S.R. and Spitler, L.G. 2004, Phys. Plasma~11, 1969
\bibitem[Spangler et al (2010)]{Spangler10} Spangler, S.R., Savage, A.H., and Redfield, S. 2010, Nonlinear Proc. Geophys.~17, 785
\bibitem[Spangler et al (2011)]{Spangler11} Spangler, S.R., Savage, A.H., and Redfield, S. 2011, arXiv 1012.4121, in ``Partially Ionized Plasmas throughout the Universe'', American Institute of Physics Conference Proceedings (V. Florinski,ed), in press 
\bibitem[Tu and Marsch (1995)]{Tu95} Tu, C.Y. and Marsch, E. 1995, \ssr~73, 1
\bibitem[Uzdensky (2007)]{Uzdensky07} Uzdensky, D. 2007, \apj~671, 2139
\bibitem[Van Eck et al (2011)]{VanEck11} Van Eck, C.L. et al 2011, \apj~728, 97 
\bibitem[Zank and Matthaeus (1992)]{Zank92} Zank, G.P. and Matthaeus, W.H. 1992, J. Plasma Phys.~48, 85
\end{thebibliography}
\end{document}